\documentclass[12pt,preprint]{aastex}

\newcommand\HII{H\,{\sc ii}~}

\newcommand\kms{km~s$^{-1}$}
\newcommand\cmsq{cm$^{-2}$~}
\newcommand\cc{cm$^{-3}$}

\newcommand\mjb{mJy~beam$^{-1}$}
\newcommand\jb{Jy~beam$^{-1}$}
\newcommand\jyb{Jy~b$^{-1}$}
\newcommand\dv{$\Delta v_{FWHM}$}
\newcommand\pp{$^{\prime\prime}$}
\newcommand\um{$\mu$m~}

\newcommand\q{$\sim$}
\newcommand\nh{$N_{\rm H}$}

\newcommand\h{H$_{2}$~}
\newcommand\msolar{M$_{\odot}$}  
\newcommand\lsun{L$_{\odot}$}

\newcommand{\methanol}{CH$_3$OH}

\newcommand{\formald}{H$_{2}$CO}
\newcommand{\hcthreen}{HC$_{3}$N}
\newcommand{\eup}{E$_{upper}$}
\newcommand{\chhhohline}{$4_{\rm +2,2,0} \rightarrow 3_{\rm +1,2,0}$} 
\newcommand{\hhcolinea}{$3_{\rm 0,3} \rightarrow 2_{\rm 0,2}$}
\newcommand{\hhcolineb}{$3_{\rm 2,2} \rightarrow 2_{\rm 2,1}$}
\newcommand{\hhcolinec}{$3_{\rm 2,1} \rightarrow 2_{\rm 2,0}$}
\newcommand{\hhcolineanoarr}{$3_{\rm 0,3}$--$2_{\rm 0,2}$}

\newcommand{\hhcoa}{$3_{\rm 0,3}$--$2_{\rm 0,2}$}
\newcommand{\hhcob}{$3_{\rm 2,2}$--$2_{\rm 2,1}$}
\newcommand{\sioline}{5$\rightarrow$4}
\newcommand{\dcnline}{3$\rightarrow$2}

\newcommand{\hcccnlinea}{24$\rightarrow$23}
\newcommand{\hcccnlineb}{25$\rightarrow$24}
\newcommand{\hcccnlineanoarr}{24--23}

\newcommand{\chhhoh}{CH$_3$OH}
\newcommand{\hcccn}{HC$_3$N}
\newcommand{\hhco}{H$_2$CO}
\newcommand{\cna}{$2_{0,3,3}\rightarrow 1_{0,2,2}$}
\newcommand{\cnb}{$2_{0,3,4}\rightarrow 1_{0,2,3}$}
\newcommand{\cnc}{$2_{0,3,2}\rightarrow 1_{0,2,1}$}
\newcommand{\cnd}{$2_{0,3,2}\rightarrow 1_{0,2,2}$} 
\newcommand{\cne}{$2_{0,3,3}\rightarrow 1_{0,2,3}$} 
\newcommand\water{H$_{2}$O}
\newcommand\ujyb{$\mu$Jy~beam$^{-1}$}
\newcommand{\smane}{SMA1-NE}
\newcommand{\smasw}{SMA1-SW}
\newcommand{\meth}{CH$_3$OH}
\newcommand{\form}{H$_{2}$CO}
\newcommand{\vlsr}{$v_{LSR}$}

\begin{document}
\shortauthors{Cyganowski et al.}

\title{Evidence for a Massive Protocluster in S255N}
\author{C.J. Cyganowski\altaffilmark{1}, C.L. Brogan\altaffilmark{2}, 
T.R. Hunter\altaffilmark{2}}

\email{ccyganow@astro.wisc.edu}
\email{cbrogan@nrao.edu}
\email{thunter@nrao.edu}

\altaffiltext{1}{University of Wisconsin, Madison, WI 53706}
\altaffiltext{2}{NRAO, 520 Edgemont Rd, Charlottesville, VA 22903}

\begin{abstract}

S255N is a luminous far-infrared source that contains many indications
of active star formation but lacks a prominent near-infrared stellar
cluster.  We present mid-infrared through radio observations aimed at
exploring the evolutionary state of this region.  Our observations
include 1.3~mm continuum and spectral line data from the Submillimeter
Array, Very Large Array 3.6~cm continuum and 1.3~cm water maser data,
and multicolor IRAC images from the {\em Spitzer Space Telescope}.
The cometary morphology of the previously-known UC\HII\ region
G192.584-0.041 is clearly revealed in our sensitive,
multi-configuration 3.6~cm images.  The 1.3~mm continuum emission has
been resolved into three compact cores, all of which are dominated by
dust emission and have radii $< 7000$ AU.  The mass estimates for
these cores range from 6 to 35 M$_\sun$.  The centroid of the
brightest dust core (SMA1) is offset by $1.1''$ (2800 AU) from the
peak of the cometary UC\HII\ region and exhibits the strongest
HC$_3$N, CN, and DCN line emission in the region.  SMA1 also exhibits
compact CH$_3$OH, SiO, and H$_2$CO emission and likely contains a
young hot core. We find spatial and kinematic evidence that SMA1 may
contain further multiplicity, with one of the components coincident
with a newly-detected H$_2$O maser.  There are no mid-infrared point
source counterparts to any of the dust cores, further suggesting an
early evolutionary phase for these objects.  The dominant mid-infrared
emission is a diffuse, broadband component that traces the surface of
the cometary UC\HII\ region but is obscured by foreground material on
its southern edge.  An additional 4.5~$\mu$m linear feature emanating
to the northeast of SMA1 is aligned with a cluster of methanol masers
and likely traces a outflow from a protostar within SMA1.  Our
observations provide direct evidence that S255N is forming a cluster
of intermediate to high-mass stars.

\end{abstract}


\section{Introduction}

The formation and early evolution of stellar clusters occurs in deeply
embedded regions of giant molecular clouds \citep{Lada03}.  While much
has been learned from recent surveys in the infrared
\citep{Gutermuth05,Muench02}, the earliest stages of cluster formation
will (at least in many cases) be hidden from all but the longest
IR/millimeter wavelengths.  Due to the small size scales of young
clusters--multiplicity of protostars has been observed on scales of a
few thousand AU \citep[e.g.][]{Megeath05}--high angular resolution is
necessary to resolve individual objects, particularly in massive star
forming regions which lie at relatively large distances ($>1$ kpc).
These considerations point to millimeter-wavelength interferometric
observations of thermal dust continuum emission as an effective means
of searching for clusters of young protostars, as the mm emission
remains optically thin up to high column densities
(\nh\q10$^{25}$~\cmsq).

\begin{figure}
\epsscale{1.0}
\plotone{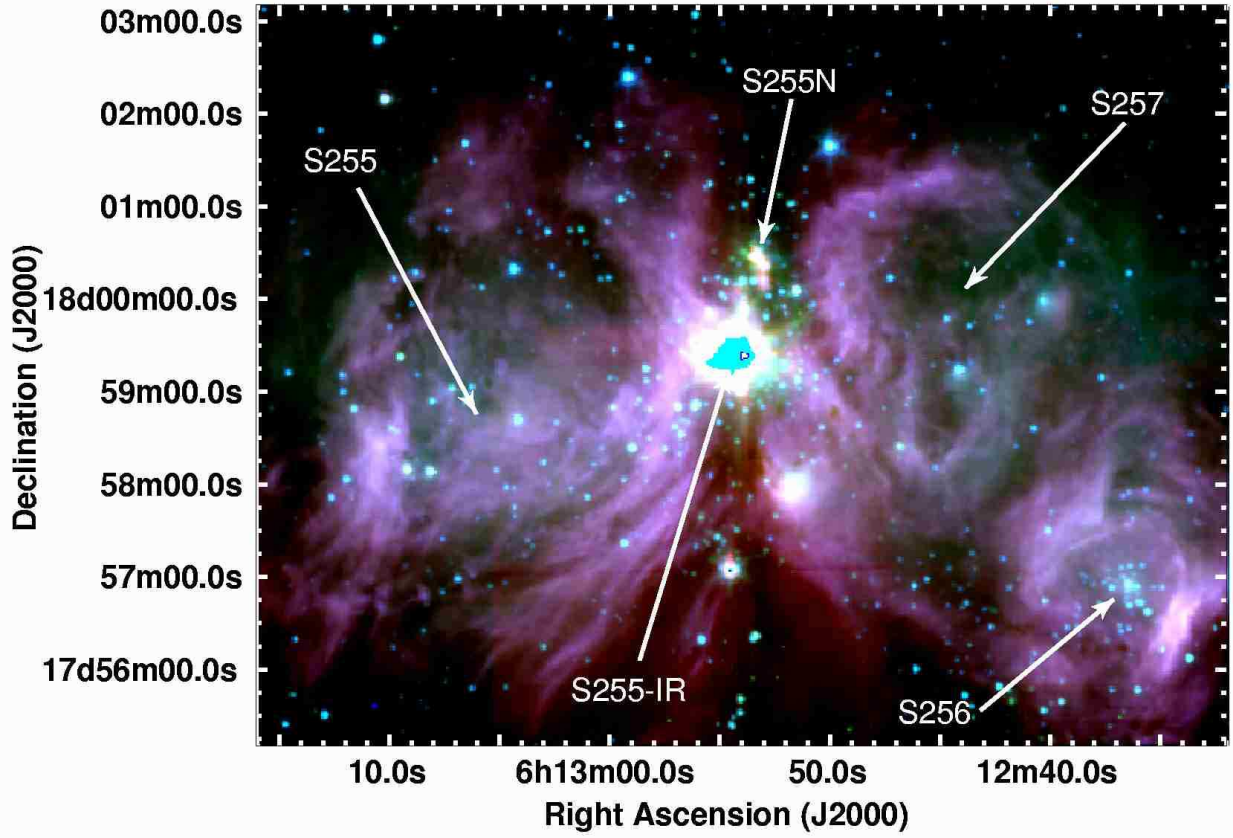}
\caption{Three-color {\em Spitzer} IRAC image of S255N and its
  surroundings showing 8.0 \um (red), 4.5~\um (green), and 3.6 \um
  (blue). S255N lies in a complex region of past and ongoing massive
  star-formation. }
\label{irbig}
\end{figure}

Located $\sim 1'$ north of the luminous infrared cluster S255IR, S255N
is a promising target in the search for young protoclusters.  As
illustrated in Figure~\ref{irbig}, S255N is located in a complicated
region of past and ongoing massive star formation.  S255N and S255IR
(saturated in this mid-IR {\em Spitzer} IRAC image) lie between two
large \HII regions, S255 and S257.  Large-scale $^{12}$CO and HCN
observations show that S255IR and S255N occupy opposite ends of a
molecular ridge between the \HII regions \citep{Heyer89}.  The total
luminosity of S255N ($\sim 1\times 10^5$~\lsun\/) is about twice that
of S255IR \citep{Minier05}, and single-dish continuum and spectral
line observations at submillimeter and millimeter wavelengths have
established the presence of large column densities of dust and gas
toward both regions \citep[e.g.][]{Richardson85, Mezger88,
Zinchenko97, Minier05}.  Observations at infrared and radio
wavelengths, however, suggest that S255N is the younger of the two
regions.  For example, S255IR is bright ($>70$~Jy) in all infrared
bands of the \emph{Midcourse Space Experiment (MSX)} and contains a
well-developed near-IR cluster of early-B type stars
\citep[S255-2][]{Howard97,Itoh01}, a cluster of compact \HII regions
\citep{Snell86}, and a wealth of complex \h emission features
\citep{Miralles97}.  In contrast, S255N (also called S255 FIR1 and
G192.60-MM1) is undetected in \emph{MSX} images at wavelengths shorter
than 21~\um \citep{Crowther03}, and contains only a single cometary UC
\HII region, G192.584-0.041 \citep[e.g.][]{Kurtz94}.  

Additional evidence for protostellar activity in S255N exists in the
form of outflow tracers.  For example, two small knots of \h emission
bracket the UC \HII region and may be tracing an outflow
\citep{Miralles97}.  In beamsizes of $\sim 50''$, redshifted CO
emission \citep{Heyer89} and highly-blueshifted OH absorption
\citep{Ruiz92} are also seen toward the UC \HII region.  Finally,
44~GHz (7$_{0}$-6$_{1}$) A$^{+}$ Class I methanol masers form a linear
feature extending northeast of the UC \HII region \citep{Kurtz04};
masers of this type have been observed in association with molecular
outflows in other objects \citep{Plambeck90,Kurtz04}.

At this stage, further understanding of the state of star formation in
S255N requires resolving the mm dust continuum emission in order to
search for additional compact sources that may be present in the
vicinity of the UC \HII region.  The high angular resolution now
available with the Submillimeter Array (SMA)\footnote{The
Submillimeter Array is a joint project between the Smithsonian
Astrophysical Observatory and the Academia Sinica Institute of
Astronomy and Astrophysics and is funded by the Smithsonian
Institution and the Academia Sinica.} makes this goal possible, while
the wide bandwidth allows simultaneous observation of many spectral
lines, which are sensitive to a range of gas conditions across the
region.  We describe our observations in section 2, present our
results in section 3, and discuss our interpretations in section 4.

A range of distances to S255 and S255N have been used in the
literature.  At the extremes, \citet{Georgelin73} find an optical
distance of 3.2~kpc to each of the adjacent \HII regions S254 and S257
(located west of S255N), while \citet{HM90} find a distance of 1.1~kpc
to the optical \HII region S255 (located east of S255N, see
Figure~\ref{irbig}) based on spectroscopy of its exciting star.  Using
a LSR velocity of +9~\kms\/ (typical of the centroid velocities of our
observed lines), and a Galactic center distance of 8.5 kpc, we find a
kinematic distance to S255N of 2.2~kpc using the rotation curve of
\citet{BB93}. In this paper we adopt a distance of 2.6 kpc for S255N
\citep[also see][]{Moffat1979,Minier05}.

\section{Observations}

\subsection{Submillimeter Array (SMA)}

Our SMA observations toward S255N were obtained on December 06, 2003
in the compact configuration.  Six antennas were operational and the
projected baseline lengths ranged from 12 to 62~k$\lambda$, resulting
in a synthesized beam of $4\farcs7 \times 2\farcs4$
(P.A.=$-45.65\arcdeg$).  In this configuration, the interferometer is
not sensitive to smooth structures larger than 17\pp.  The phase
center was $06^{\rm h}12^{\rm m}53^{\rm s}.56$, $18^{\circ}00'25''.0$
(J2000), and the double-sideband SIS receivers were tuned to an LO
frequency of 222.78~GHz. With a bandwidth of \q2~GHz, the correlator
covered 216.796-218.764~GHz in lower sideband (LSB) and
226.796-228.764~GHz in upper sideband (USB).  The data were resampled
to provide uniform spectral resolution of 1.12~\kms.  The zenith
opacity as measured by the 225~GHz tipping radiometer at the Caltech
Submillimeter Observatory (CSO) was 0.18 at the beginning of the
observations and fell as low as 0.16. Data recorded late in the
observing session, when the opacity had climbed to 0.3, were not used.
The typical system temperature at source transit was 200~K.  The
primary beamsize of the 6-meter SMA antennas at this frequency is
56\pp.

Initial calibration of the data was performed in Miriad. The gain
calibrators were J0739+016 and J0423-013.  J0423-013 was also used for
passband calibration.  Subsequent processing was carried out in AIPS.
The continuum emission was measured using line-free channels and
removed in the $u-v$ plane. The resulting continuum-only data were
then self-calibrated; these complex gain solutions were also applied
to the continuum-subtracted line data.  The absolute position
uncertainty is estimated to be $0.''3$ and the amplitude calibration
is accurate to 20\%. For maximum sensitivity, both the continuum and
line data were imaged with natural weighting.  The rms sensitivity of
the continuum image is 4~\mjb ~(8 mK) and the rms sensitivity of the
line data is 89~\mjb ~(170 mK).

\subsection{Very Large Array (VLA)}

Archival 3.6~cm data from the NRAO\footnote{The National Radio
Astronomy Observatory is a facility of the National Science Foundation
operated under agreement by the Associated Universities, Inc.} Very
Large Array (VLA) were calibrated and imaged in AIPS.  The observation
date was 2003 June 15 and the total time on source was
$\sim$70~minutes (project code AH819).  The VLA was in
A-configuration, in which the interferometer is not sensitive to
smooth structures larger than \q7\pp.  The bandwidth was 50~MHz in two
IFs.  The flux calibrator was 3C147, and the gain calibrator was
J0613+131.  The synthesized beam of the 3.6~cm continuum image is
$0\farcs27 \times 0\farcs23$ (P.A.=$77.51\arcdeg$) and the rms
sensitivity is 18~$\mu$Jy~beam$^{-1}$. Archival 3.6~cm data from
the B configuration (in which the interferometer is not sensitive to
smooth structures larger than \q20\pp), observed on 1990 August 27
(project code AM301), and from the the CnB configuration, observed on
2005 June 21 (project code TSTCJC), were also reduced and combined
with the A-configuration data.  The resulting image, made with a UV
taper at 500~k$\lambda$, has a synthesized beam of $1\farcs03 \times
0\farcs84$ (P.A.=$-81.42\arcdeg$) and a rms sensitivity of
34~$\mu$Jy~beam$^{-1}$.

Archival 1.3~cm data from the VLA A-configuration (project code AC299)
were analyzed for water maser emission at 22.235~GHz.  The observation
date was 1991 August 1, total on-source integration time was $\sim$10
minutes. The phase center of this observation was toward the IR
cluster $\sim 1\arcmin$ to the south of S255N, hence a correction for
the primary beam attenuation has been applied to the data. The
bandpass calibrator was 3C84 and the flux calibration was derived
assuming a flux density of 2.16~Jy for J0528+134.  The synthesized
beam is $0\farcs18 \times 0\farcs16$ (P.A.=$-56.55\arcdeg$), the
spectral line channel spacing is 0.33~\kms\/, and the rms noise is
0.13 \jb\/.

\subsection{Spitzer Space Telescope}

Mid-infrared images of S255 were obtained with the IRAC camera
\citep{Fazio04} on the {\em Spitzer Space Telescope} as part of
Guaranteed Time Observations program 201 (P.I. G. Fazio) on 12 March
2004.  Integrations of 0.4 s and 10.4 s were taken in the high dynamic
range mode; S255N is not saturated in the longer exposures, and only
the 10.4 s exposures are discussed in this paper.  Four 10.4 s exposures
covered S255N, for a total integration time on S255N of 41.6 s.
Mosaiced post-BCD 3.6, 4.5, 5.8, and 8.0 \um images, calibrated and
processed using pipeline version S13.2.0, were downloaded from the
{\em Spitzer} data archive.

\subsection{Caltech Submillimeter Observatory}

Our submillimeter continuum observations were obtained at the CSO
using the Submillimeter High Angular Resolution Camera (SHARC), a
$^{3}$He-cooled monolithic silicon bolometer array of 24 pixels in a
linear arrangement \citep{Hunter96,Wang96}.  For a typical dust source
with a submillimeter spectral index of $\sim4$, the effective
frequency of the broadband 350~$\mu$m filter is 852~GHz and the
bandwidth is 103~GHz.  An on-the-fly (OTF) map of S255 was obtained on
21 December 1995 by scanning the array through the source in azimuth
while the secondary mirror was chopping at a rate of 4.1 Hz and a
throw of $88''$.  Successive scans were made after stepping the array
in elevation by increments of $5''$.  Airmass corrections were applied
to each scan using the opacity derived from frequent scans of Saturn
during the night.  The map data were restored with a NOD2 dual beam
restoration algorithm \citep{Emerson79} and transformed into
equatorial coordinates. The resulting image was smoothed with a
Gaussian to produce an effective half-power beamsize of $15''$.

\section{Results}

\subsection{Continuum emission}

\begin{figure}
\epsscale{0.5}
\plotone{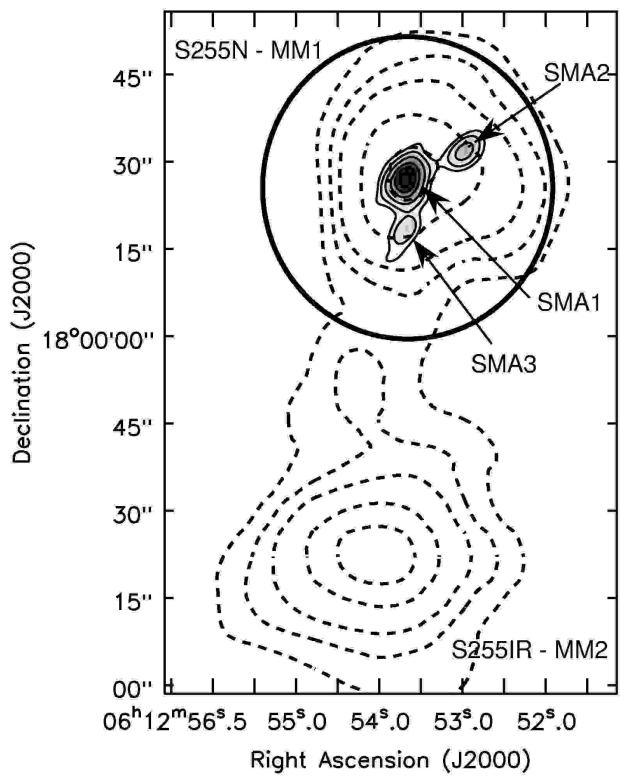}
\caption{Greyscale and solid contours of the SMA 1.3~mm continuum with
  dotted contours of the CSO 350 \um continuum superposed.  The primary
  beam of the SMA (56\pp~ at 218.7 GHz) is indicated with a black
  circle.  Naming conventions for mm and submm sources used in the
  literature and in this paper are also indicated.  The black 1.3~mm
  contour levels are (-3, 3, 7, 15, 31, 47, 63) $\times$ 4 \mjb ~(the rms
  noise), observed with a $4\farcs7\times
  2\farcs4$ (P.A.=$-45.65\arcdeg$) beam.  The dotted 350 \um contour
  levels are (2, 2.5, 3, 4, 5) $\times$ 16.5 Jy beam$^{-1}$,
  resolution 15\pp.}
\label{csocont}
\end{figure}

Our 1.3~mm SMA continuum data resolve three distinct sources within
the previously-observed submm/mm clump of S255N \citep[aka S255 FIR1
and G192.60-MM1:][]{Jaffe84, Mezger88, Minier05}.  Figure
~\ref{csocont} shows the SMA 1.3~mm continuum image with CSO 350 \um
contours superposed. As illustrated in Figure~\ref{csocont}, the
strongest SMA 1.3~mm emission peak coincides with the CSO 350 \um peak
(resolution $15\arcsec$).  The CSO 350 \um\/ integrated flux density
is 575$\pm 20$ Jy, consistent to within 10\% of the value predicted by
the dust spectral energy distribution (SED) models of
\citet{Minier05}.  The three mm sources resolved with the SMA are
designated SMA1, SMA2, and SMA3 in order of descending peak intensity.
The observed properties of each source (peak intensity, brightness
temperature, integrated flux density, and source size) are listed in
Table~\ref{cont}, and the sources are labeled in Figure~\ref{csocont}.
The integrated flux densities and source sizes listed in
Table~\ref{cont} were determined by fitting a single Gaussian
component to each SMA source.  SMA1 was not well fit by a single
Gaussian, indicating that the observed continuum emission may arise
from multiple sources unresolved by the SMA beam; this issue is
discussed further in \S4.2.  The total flux density of the three
compact mm sources is 0.79$\pm$0.16~Jy; stated errors include the 20\%
uncertainty in flux calibration.  This total corresponds to 15$\pm$3\%
of the single-dish flux density measured by \citet{Minier05} with the
SEST 15m telescope at 1.2mm (resolution 24\pp).

\begin{figure}
\plotone{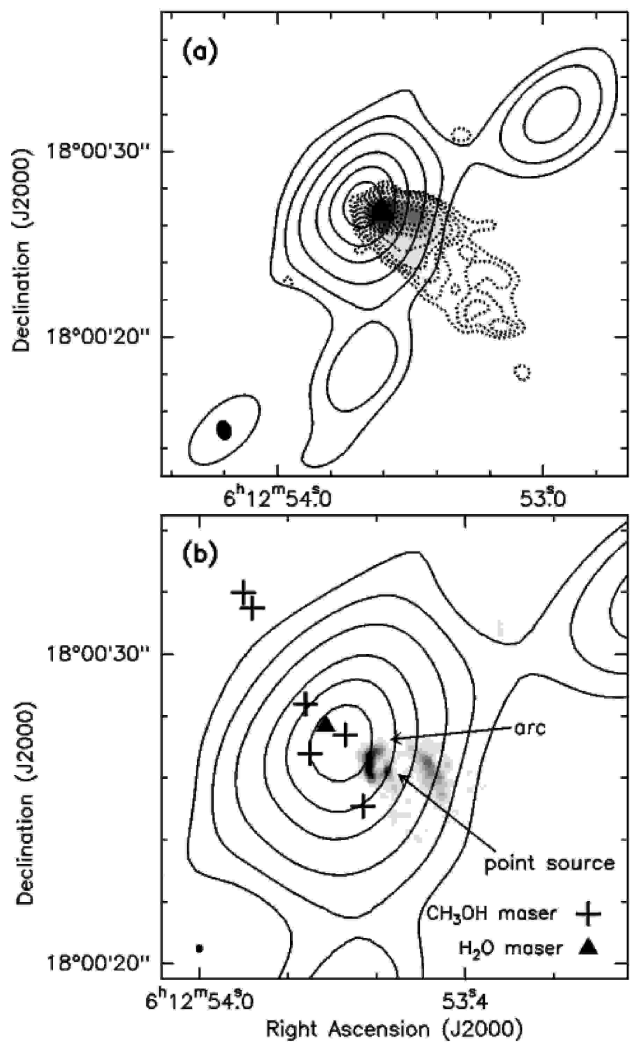}
\caption{(a) Black contours of the SMA 1.3~mm continuum, observed with a
  $4\farcs7\times 2\farcs4$ (P.A.=$-45.65\arcdeg$) beam, with greyscale
  and dotted contours of the 3.6~cm continuum, observed with an
  $1\farcs03\times 0\farcs84$ beam (P.A.=$-81.42\arcdeg$), superposed.
  The black 1.3~mm contour levels are (-3, 3, 7, 15, 31, 47,
  63) $\times$ 4 \mjb~ (the rms noise).  The dotted 3.6~cm contour levels
  are (-3, 3, 5, 7, 11, 21, 41, 61, 81, 101, 121, 161)~$\times$~34~$
  \mu$Jy~beam$^{-1}$ (the rms noise). The SMA beam (black ellipse) and
  VLA beam (filled black ellipse) are plotted at lower left. (b) Black
  contours of the SMA 1.3~mm continuum with greyscale of
  high-resolution ($0\farcs27\times 0\farcs23$, P.A.=$77.51\arcdeg$
  beam) 3.6~cm continuum superposed.  The positions of methanol masers
  and of the newly-reported water maser are marked.  The VLA beam
  (filled black ellipse) is plotted at lower left.}
\label{masers}
\end{figure}

Figure~\ref{masers} compares the morphology of the 1.3~mm dust
continuum emission with the 3.6~cm free-free continuum emission from
the cometary UC \HII region, G192.584-0.041.  Figure~\ref{masers}a
shows the lower-resolution ($1\farcs03 \times 0\farcs84$) 3.6~cm VLA
image superposed on the SMA 1.3~mm continuum, while
Figure~\ref{masers}b shows the high-resolution VLA 3.6~cm image
($0\farcs27 \times 0\farcs23$), with the positions of the
newly-reported water maser (see \S3.2) and the Class I methanol masers
detected by \citet{Kurtz04} indicated.

The $\sim 1\arcsec$ resolution 3.6~cm VLA image presented in
Figure~\ref{masers}a provides the most detailed view to date of the
diffuse cometary ``tail'' of the UC \HII region. The integrated flux
density of G192.584-0.041 measured from this image is 26.0$\pm$0.1
mJy.  Based on our measurement and published 2~cm integrated flux
densities for G192.584-0.041 \citep{Kurtz94,Rengarajan96}, the
spectral index from 3.6~cm to 2~cm is \q-0.1, consistent with
optically thin free-free emission.  The flux density is consistent
with a single exciting star of spectral type B0.5 \citep[as determined
by][]{Kurtz94,Snell86}.  Extrapolating to 1.3~mm, we estimate the
free-free contribution of G192.584-0.041 to the 1.3~mm flux density of
SMA1 to be $\lesssim$20 mJy ($\leq$3.5\%).

The 3.6~cm VLA image presented in Figure~\ref{masers}b is the
highest-resolution cm-wavelength image of G192.584-0.041 to date.
With a resolution of $0\farcs27 \times 0\farcs23$, the continuum
emission from G192.584-0.041 is resolved into three components (east
to west): an arc, a point source, and an extended feature (the
extended ``tail'' is resolved out in the higher resolution data).  All
three of these components overlap with the eastern side of SMA1, but
none is coincident with the mm emission peak, in agreement with the
estimate that the free-free contribution at 1.3~mm is quite small.
The arc, which is oriented with its convex side towards the mm peak,
contains the brightest 3.6~cm emission.  The 3.6~cm peak is located in
the southern part of the arc, east of the point source, and is offset
by 1\farcs1 (\q2800~AU) from the location of the SMA1 mm continuum
peak determined by fitting a single Gaussian component.  The 3.6~cm
point source, which is located west of the arc and faces its concave
side, is offset by 1\farcs6 (\q4,200~AU) from the SMA1 mm continuum
peak.  The peak brightness temperature of the 3.6~cm point source is
only 122 K at the current resolution.  Absent a second radio frequency
image with comparable resolution, it is not currently possible to
ascertain the spectral indices of the individual components.  The
3.6~cm images presented in Figure~\ref{masers}(a-b) place strong
limits on the presence of any additional \HII regions in S255N.  Other
than G192.584-0.041, no cm-wavelength emission is detected down to a
5$\sigma$ limit of 90~\ujyb ~(high-resolution image).

\subsection{Water maser emission}

Water maser emission was detected at the position $06^{\rm h}12^{\rm
m}53^{\rm s}.71$, $18^{\circ}00'27\farcs6$ (J2000), offset $0\farcs9$
(\q2,300~AU at 2.6~kpc) to the northeast of the 1.3~mm continuum
emission peak of SMA1, as determined by fitting a single Gaussian
component.  This is the first report of water maser emission from
S255N.  The peak intensity is 2.8~Jy~beam$^{-1}$ (corrected for
primary beam attenuation) at $v_{LSR}$=+8.5 \kms. The line is barely
resolved by the 0.33 \kms\ spectral resolution.

The positions of the Class I 44~GHz (7$_{0}$-6$_{1}$) A$^{+}$ methanol
masers detected by \citet{Kurtz04} are marked with crosses in
Fig.~\ref{masers}b, which shows the 1.3~mm and 3.6~cm continuum emission
and the newly-reported water maser.  \citet{Kurtz04} estimate an
astrometric uncertainty of $0\farcs5$ for the \methanol\/ maser spots,
while the absolute astrometry of the \water\/ maser is better than
0\farcs1.  The position of one of the \methanol\/ maser spots is
consistent with SMA1, within the astrometric uncertainty.  The newly
detected water maser is $<$ 1\arcsec ~from two \methanol\/ masers, and
falls into the linear pattern of 44~GHz (7$_{0}$-6$_{1}$) A$^{+}$
\methanol\ maser spots that extends northeast from SMA1.

\subsection{Line emission}

\begin{figure}
\includegraphics[angle=-90,width=6in]{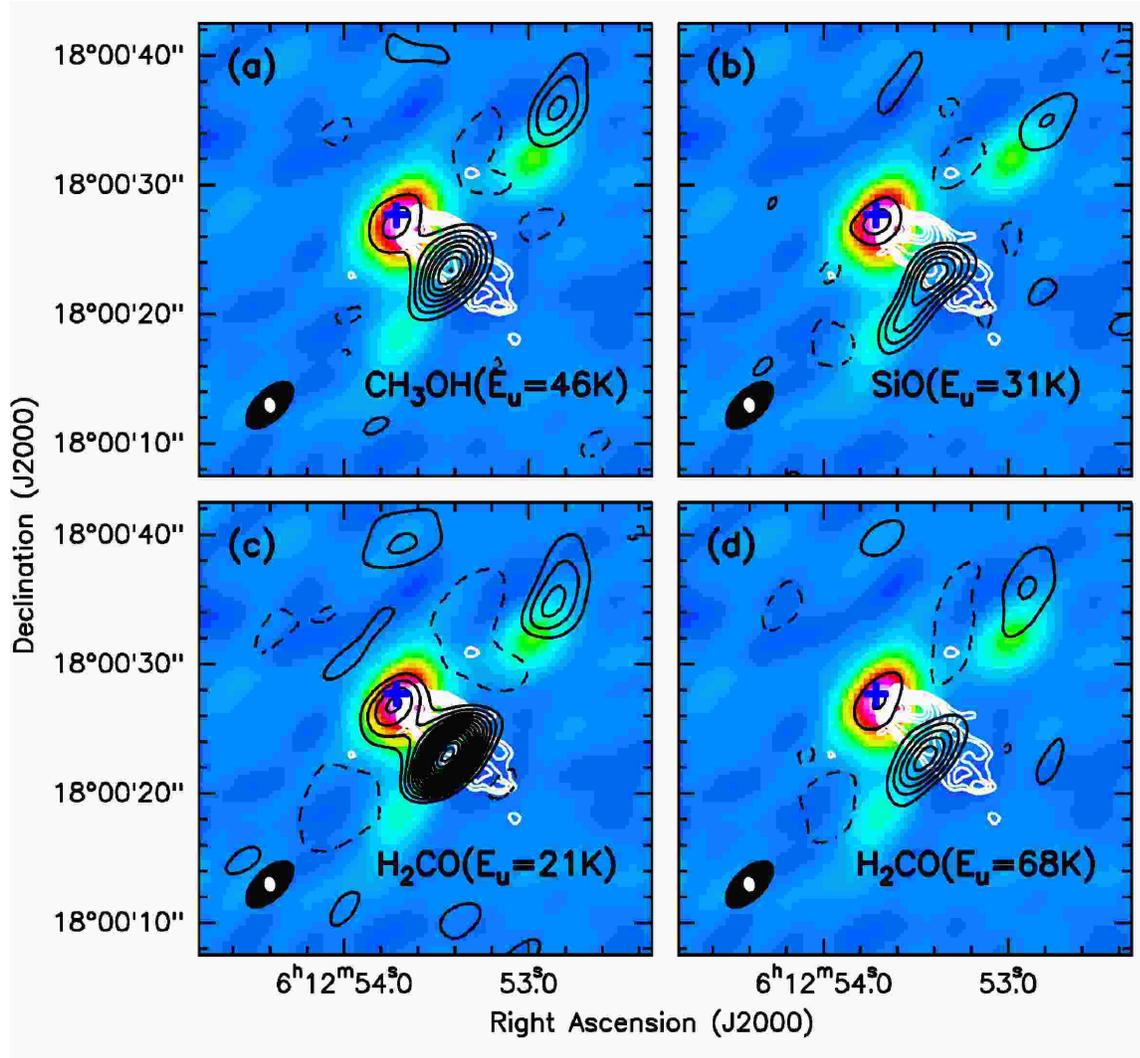}
\caption{(a-d) Colorscale of the 1.3~mm continuum with black integrated
intensity contours for molecules that exhibit emission from multiple
positions. The molecular species and upper state energies are
indicated in the lower right of each panel (also see Table 2).  The
white contours show 3.6~cm emission from the UC \HII region
G192.584-0.041. The blue cross marks the position of the newly
detected water maser.  The black integrated intensity contour levels
are (-3, 3, 5, 7, 9, 11, 13, 15, 17, 19, 21, 23, 25, 27, 29) $\times$
the rms noise levels: (a) \methanol~ 0.82 \jb*\kms, (b) SiO 1.35
~\jb*\kms, (c) \formald~\hhcolinea~ 0.8 \jb*\kms, (d)
\formald~\hhcolineb~ 0.76 \jb*\kms.  The white 3.6~cm contour levels
are (-3, 3, 5, 7, 11, 21, 41, 61, 81, 101, 121, 161)~$\times$~34~$
\mu$Jy~beam$^{-1}$ (the rms noise). The $4\farcs7\times
2\farcs4$ (P.A.=$-45.65\arcdeg$) SMA beam (filled black ellipse) and
$1\farcs03\times 0\farcs84$ (P.A.=$-81.42\arcdeg$) VLA beam (filled white
ellipse) are shown at lower left in each panel.}
\label{emom}
\end{figure}

\begin{figure}
\includegraphics[angle=-90,width=6in]{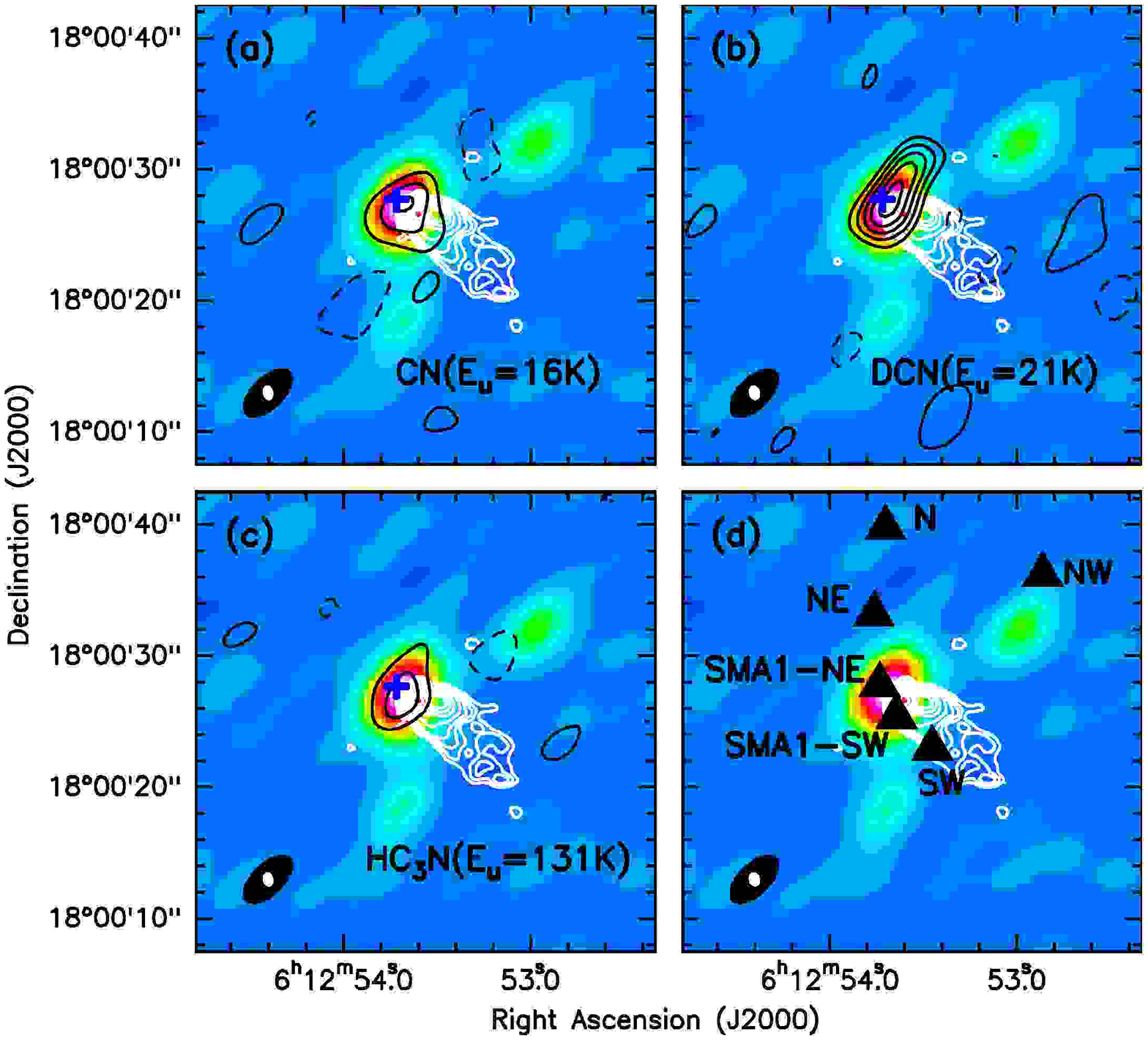}
\caption{(a-c) Similar to Figure~\ref{emom}a-d except
integrated intensity images are shown for molecules that are only
detected in the vicinity of SMA1. The black integrated intensity
contour levels are (-3, 3, 5, 7, 9, 11) $\times$ the rms noise
levels: (a) CN 0.9 \jb*\kms, (b) DCN 0.55 \jb*\kms, and (c)
\hcccn~\hcccnlinea~0.68 \jb*\kms.  (d) Finding chart for
representative line profiles shown in
Figures~\ref{epro} \& ~\ref{cpro} and 
discussed in \S3.3.}
\label{cmom}
\end{figure}

Molecular line emission from \formald\/, \methanol\/, SiO, CN, DCN,
and \hcccn\/ is detected in S255N; the specific transitions,
frequencies, and upper state energies are listed in Table~\ref{trans}.
The lines detected in S255N are the same as those detected in the
spectral regions covered by our sidebands by \citet{Sutton85} in their
line survey of the Orion A molecular cloud, which is similar to our
data in spectral resolution (1.3 km/s), rms sensitivity (0.2~K), and
linear size scale (30\pp = \q13,500~AU at 450 pc). Integrated
intensity images for \methanol\/, SiO, and \formald\/ are presented in
Figure~\ref{emom}(a-d) and for DCN, \hcccn\/, and CN in
Figure~\ref{cmom}(a-c).  The distributions of molecular emission
observed in S255N fall into two main categories: \methanol\/,
\formald\/ and SiO exhibit emission from multiple locations, while DCN
and \hcccn\/ are detected only in the vicinity of SMA1.  CN exhibits
compact emission towards SMA1, and is also weakly detected toward
SMA3.  The CN lines have the lowest \eup~ of the observed transitions,
and the CN images show artifacts from large-scale emission resolved
out by the interferometer, suggesting that much of the CN emission
originates in an extended, cool envelope around the compact continuum
sources.

\begin{figure}
\epsscale{1.0}
\plotone{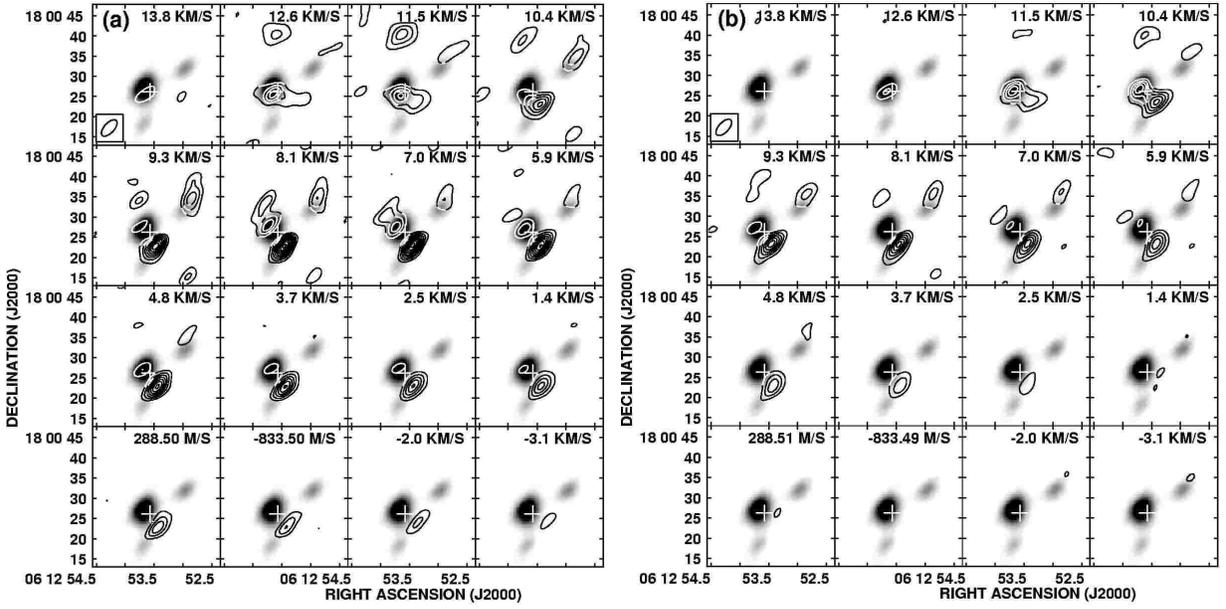}
\caption{Channel maps of (a) \formald\/(\hhcolineanoarr\/) and (b)
\methanol\/, showing line emission (contours) overlaid on the
1.3~mm continuum (greyscale).  The contours levels are (1, 2, 3, 4, 5, 6,
7, 8, 9)~$\times$ 0.28 \jb.  Each panel is labeled with the channel
velocity.  The position of the 3.6~cm point source is marked with a
white cross.}
\label{chmap}
\end{figure}

\begin{figure}
\epsscale{1.0}
\plotone{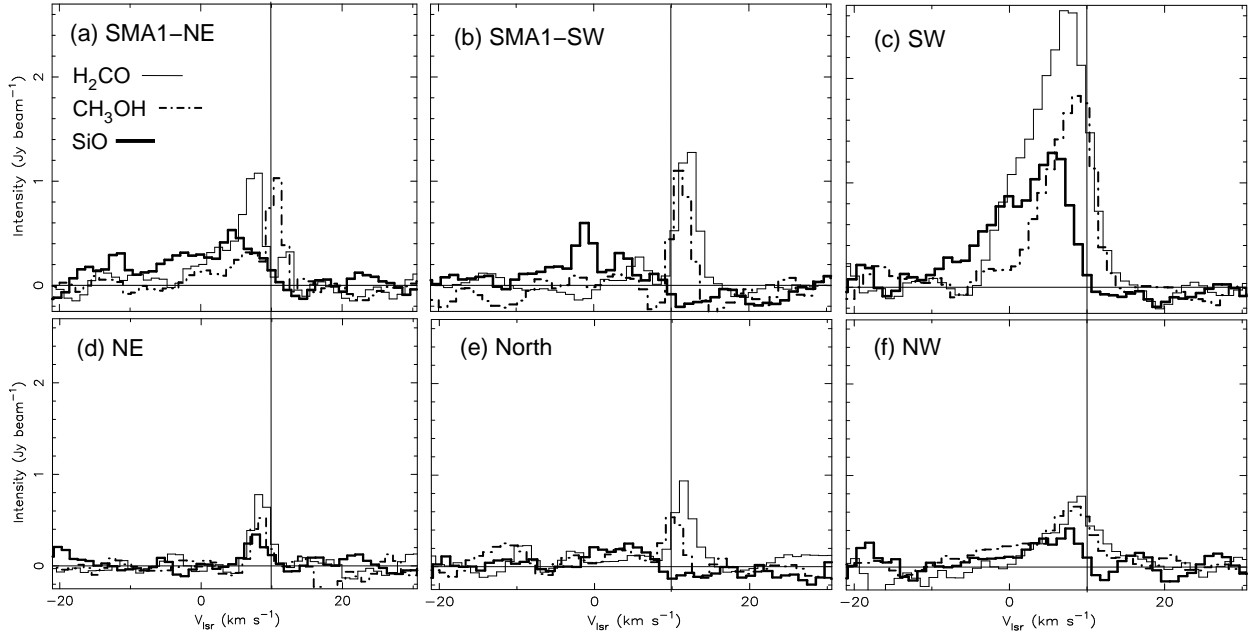}
\caption{Representative line profiles for molecules that exhibit
  emission from multiple positions, demonstrating the wide range of
  velocity and chemical behavior observed.  The positions for which
  spectra are presented are indicated on Figure~\ref{cmom}d
  and listed in Table~\ref{linepos}. 
  A vertical line is drawn at 10 \kms\/ for reference.}
\label{epro}
\end{figure}

As shown in Figure~\ref{emom}(a-d), the spatial distributions of the
integrated emission from \formald\/, \methanol\/, and SiO are similar
to one another.  The kinematics of these molecules are complex, as
illustrated in Figures ~\ref{chmap} and ~\ref{epro}, with multiple
spatially and kinematically distinct components apparent. A finder
chart for the positions of the profiles displayed in Fig.~\ref{epro}
(named after their relative positions with respect to SMA1) is shown
in Figure~\ref{cmom}(d).  The positions are listed in Table~\ref{linepos}. 
Line centroid velocities, \dv, and
integrated line intensities obtained from Gaussian fits to the
line profiles at these positions are listed in Table ~\ref{linefits}.
Unless otherwise noted, the fit parameters in Table ~\ref{linefits}
are for the strongest component in the spectrum.  Fits to SiO line
profiles are not included because the SiO line shapes are so complex.

The strongest molecular emission in S255N lies toward the ``SW''
position $\sim 6\arcsec$ to the southwest of SMA1 at a peak velocity
between $\sim 6 - 8$ \kms\/ (Figs.~\ref{emom}, \ref{chmap},
\ref{epro}c, Table~\ref{linefits}).  This molecular emission is not
coincident with any mm continuum emission and overlaps the southern
edge of the extended ``tail'' of the UC \HII region
(Fig.~\ref{emom}). The line emission toward the ``SW'' position is
broader than at any other location in S255N, with $\Delta
v_{FWHM}\sim$ 7 and 9 \kms\/ for \methanol\/ and \formald\/,
respectively, and shows pronounced blue wings.  The velocity of the
\formald\/ peak is slightly more blueshifted than \methanol\/,
while SiO is significantly blueshifted relative to both \formald\/ and
\methanol\/ (Fig.~\ref{epro}c, Table~\ref{linefits}).

\begin{figure}
\epsscale{1.0}
\plotone{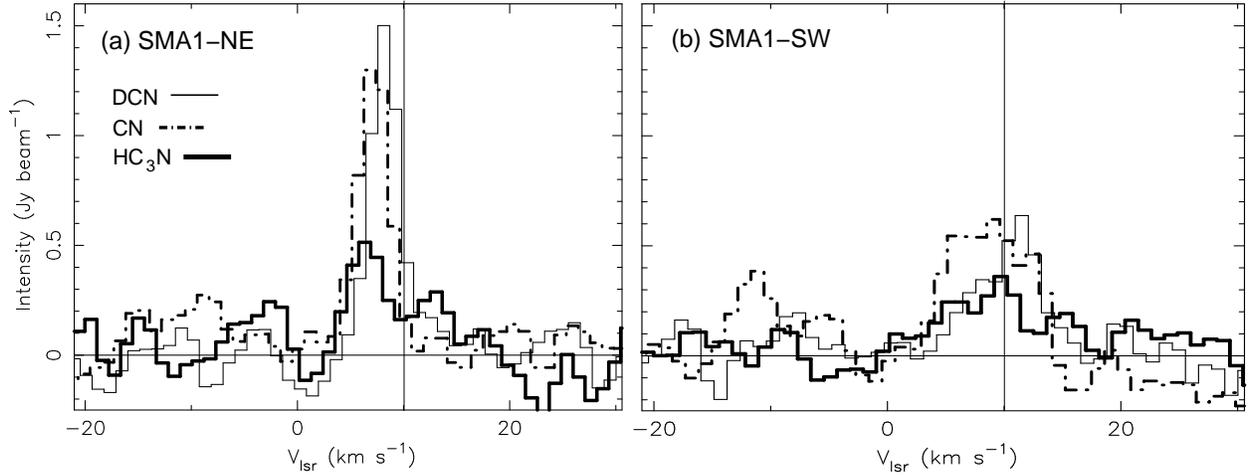}
\caption{Representative line profiles for molecules that have their
  strongest emission towards SMA1.  The positions for which spectra
  are presented are indicated on Figure~\ref{cmom}d 
  and listed in Table~\ref{linepos}. 
  A vertical line
  is drawn at 10 \kms\/ for reference.  Note that the vertical scale
  is not the same as in Figure~\ref{epro}.  For CN, the weaker
  features offset by -16 and -22 \kms\/ from the main feature are due
  to the hyperfine components (see Table~\ref{trans}).
}
\label{cpro}
\end{figure}

\begin{figure}
\epsscale{1.0}
\plotone{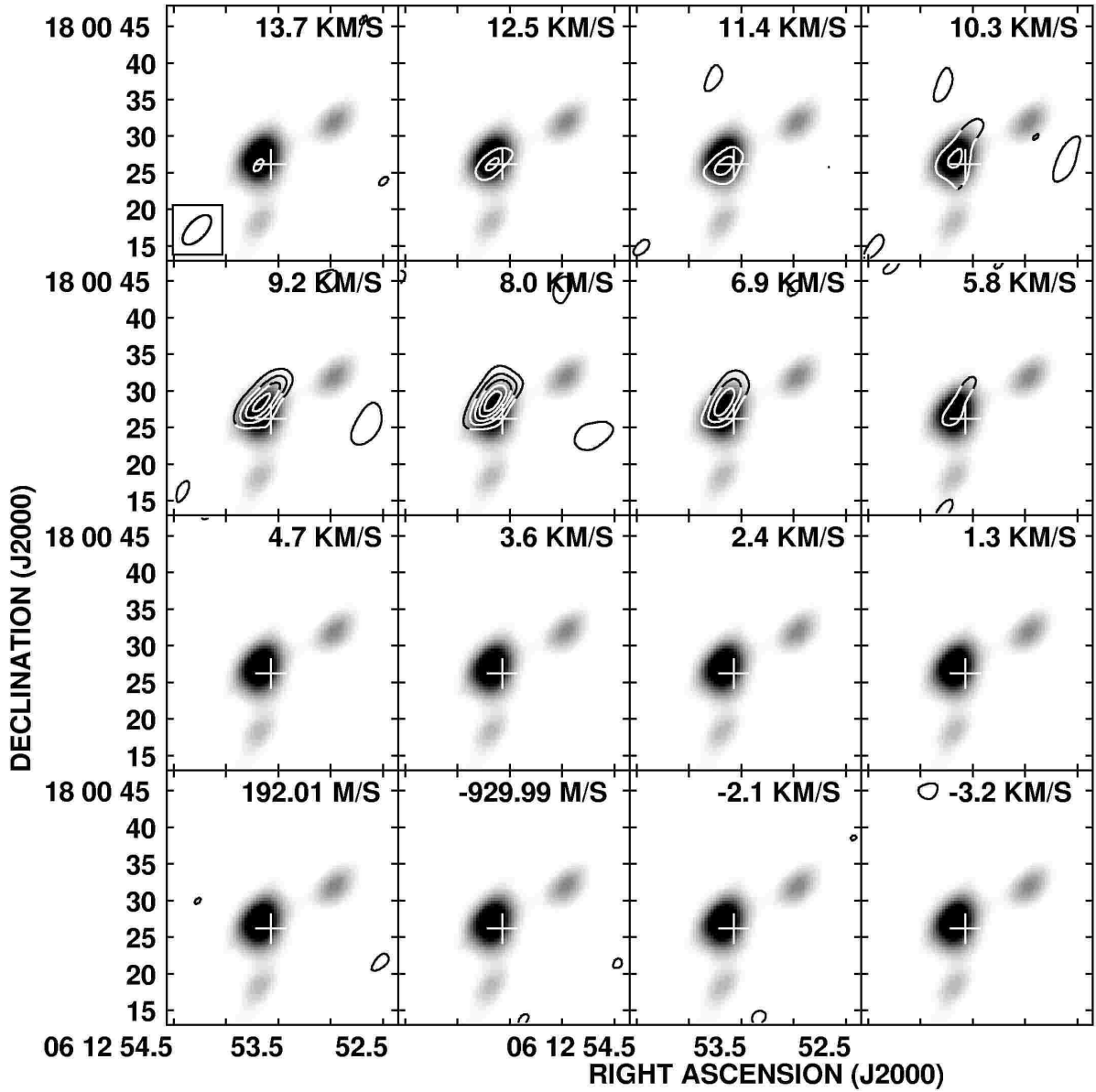}
\caption{Channel maps of DCN emission (contours) overlaid on the 1.3
  mm continuum (greyscale).   Each panel is labeled with the channel
  velocity.  The position of the 3.6~cm point source is marked with a
  white cross.}  
\label{chmapdcn}
\end{figure}

The second-brightest region of \form\/, \meth\/, and SiO emission in
S255N is located in the vicinity of SMA1, east of the cometary head of
the UC\HII\/ region (Fig. ~\ref{emom}).  DCN, CN, and \hcccn\/ have
their strongest emission in this area (Fig. ~\ref{cmom}).  Spectral
line profiles for DCN, CN, and \hcccn\/ are shown in Figure~\ref{cpro}
and channel maps for DCN are shown in Figure~\ref{chmapdcn}.  Two
positionally and kinematically distinct components are evident in
\form\/, DCN, CN, \hcccn\/, and weakly, \methanol\/, in the vicinity
of SMA1. One component (denoted \smane\/) lies $1\farcs21$ to the
northeast of the SMA1 mm peak at a velocity of $\sim 7$ \kms\/, and
the other (denoted \smasw\/) lies $1\farcs17$ southwest of the SMA1 mm
peak at $\sim 11.5$ \kms\/ (Figs.~\ref{chmap}, ~\ref{epro}a,b, and
\ref{cpro}a,b).  \smasw\/ is $1\farcs08$ east of the 3.6~cm point
source.

Interestingly, \smane\/ and \smasw\/ also show differences in their
chemical properties.  For example, \form\/ shows nearly equal strength
towards both positions, as does \hcccn, while \meth\/ is much stronger
toward \smasw\/ (Fig.~\ref{chmap}, \ref{cpro}). In contrast, DCN and
CN are both significantly stronger toward \smane\/ (Fig.~\ref{cpro}).
Some differences in the peak velocities at the two positions are also
apparent amongst species.  Relative to the other molecules, SiO is
significantly blueshifted (\vlsr$<5$ \kms\/) towards both \smasw\/ and
\smane\/ (Fig.~\ref{epro}, Table~\ref{linefits}).  DCN is slightly
redshifted relative to CN and \hcccn\/ toward both positions
(Fig.~\ref{cpro}, Table~\ref{linefits}).  The CN and \hcccn\/ lines
are also more than twice as broad as those of \form\/ or \meth\/
toward \smasw\/ (Table~\ref{linefits}).

The \form\/, \meth\/, and SiO integrated intensity peak located
\q5\arcsec~ north of the mm continuum source SMA2 (Fig.~\ref{emom},
``NW'' position in Fig.~\ref{cmom}d), is comprised of relatively weak,
broad emission (Fig.~\ref{epro}f).  The \form\/ and \meth\/ lines are
narrower than at the SW position, but broader than at any of the other
positions (Table~\ref{linefits}).  As at the other positions, the peak
of the SiO line profile is blueshifted relative to \form\/ and \meth\/
(Fig.~\ref{epro}f).

The line emission north and northeast of the SMA1 region (``N'' and
``NE'' positions, finding chart Fig.~\ref{cmom}d) consists of narrow
velocity features in \meth\/ and \form\/, and, at the NE position, SiO
(Figs.~\ref{chmap}~\&~\ref{epro}d-e).  In contrast to the other
positions, at the NE position \meth\/, \form\/, and SiO have the same
velocity, \vlsr\q8~\kms (Fig.~\ref{epro}d, Table~\ref{linefits}).  The
velocity of the \form\/ and \meth\/ emission at the N position is
similar to the \vlsr\q11.5~\kms~ component toward \smasw\/
(Table~\ref{linefits}).  The \form\/ and \meth\/ lines are narrow, and
the \meth\/ peak is slightly blueshifted relative to \form\/.
Broad and weak blueshifted SiO emission is also detected at this
position (Table~\ref{linefits}, Fig.~\ref{epro}e).

\subsection{Spitzer Space Telescope IRAC Observations}
Figure~\ref{irzoom} shows a three-color IRAC image (red 8.0 $\mu$m,
green 4.5~$\mu$m, blue 3.6 $\mu$m) of S255N, overlaid with contours of
the 1.3~mm continuum emission (yellow) and the 3.6~cm continuum
emission (white).  The positions of the Class I \methanol\/ masers
reported by \citet{Kurtz04} and the newly-reported water maser are
marked with crosses.  Most of the observed mid-IR emission is offset
to the northwest of the UC \HII region, and this diffuse emission
appears in all IRAC bands.  An exception is the linear, green 4.5~\um
emission feature that extends NE from the SMA1 mm continuum peak. No
mid-IR emission is associated with either SMA2 or SMA3, indeed these
two positions are notably absent of IR emission.

\section{Discussion}

\subsection{Mass estimates from the dust emission}

With an estimate of the dust temperature, we can estimate the masses
of the compact dust sources SMA1, SMA2, and SMA3 using a simple
isothermal model of optically thin dust emission \citep{Beltran06}:
\begin{equation}
M_{gas, thin}=\frac{R~F_{\nu}~D^2}{B(\nu,T_{d})~\kappa_{\nu}}
\end{equation}
where R is the gas-to-dust mass ratio (assumed to be 100), $F_{\nu}$
is the observed flux density, D is the distance to the source,
B($\nu,T_{d}$) is the Planck function, and $\kappa_{\nu}$ is the dust
mass opacity coefficient.  At 1.3 mm, the value of $\kappa$ for gas
densities of 10$^{6}$-10$^{8}$~\cc~does not differ much for grains
with thick or thin ice mantles; we adopt a value of $\kappa_{1.3mm}$=1
cm$^{2}$~g$^{-1}$ for all of the compact mm sources \citep{OH94}.  The
assumption of low optical depth is justified by the low observed
millimeter brightness temperatures (Table~\ref{cont}), however, for
highest accuracy we have made the small correction to our derived
masses for non-zero optical depth using the formula: $M_{\rm gas} =
M_{\rm gas,thin} \tau / (1-e^{-\tau})$.

Previous determinations of the dust temperature and mass in S255N have
relied on fitting multiple components to the (unresolved) mid-IR to mm
SED.  \citet{Minier05} fit a hot, compact core (T = 106 K, diameter =
400 AU) and an extended warm envelope (T = 44 K, diameter = 58,000AU)
to a SED comprised of \emph{MSX}, \emph{IRAS}, SCUBA, and SEST data,
assuming a distance of 2.6 kpc. The derived luminosity and gas mass
are $1.1\times 10^5$ L$_\sun$ and 220 M$_\sun$, respectively. While
many of the datapoints in the SED constructed by \citet{Minier05}
blend emission from all three SMA sources, the very compact hot core
implied by their fits would be unresolved by our SMA beam (\q12,200
$\times$ 6,200 AU at 2.6~kpc).  Thus, the hot core temperature of
106~K derived from the SED modeling provides an upper limit for the
dust temperature of the compact SMA sources.  The fitted warm envelope
temperature of 44~K is likewise a good lower limit to the temperature
of SMA1 since it dominates the 1.3~mm flux, contributing 73\% of the
total SMA flux density.

Several single dish estimates of the gas temperature are also
available. For example, Effelsberg 100-m observations of NH$_3$ (1,1)
and (2,2) (resolution $40\arcsec$) suggest that the kinetic
temperature of the gas is only $T_{kin}=23\pm 1$ K
\citep{Zinchenko97}.  Measurements of CH$_3$C$_2$H(6-5) K=0-3 toward
S255N with the Onsala 20-m (resolution $38\arcsec$) by
\citet{Malafeev2005} yield $T_{rot}=35\pm 1$ K, in better agreement
with the extended warm component derived for the dust. These authors
find significantly higher temperatures using CH$_3$C$_2$H(6-5) than
NH$_3$ towards all five sources observed (including S255) and suggest
that methyl acetylene may preferentially trace warmer/denser gas. In
any case, since beam dilution may play a significant role in these
single dish estimates, they can only provide a lower limit to the gas
temperatures on SMA sizescales.

From our SMA line data it is clear that SMA1 is the warmest of the
compact mm sources: the two detected transitions with the highest
\eup~(both \hcthreen, \eup=131~K and 142~K) are detected only towards
SMA1, suggesting this may be a hot core
\citep[e.g.][]{Hatchell98}. DCN is also seen at this warm
position. Though DCN is formed in cold clouds, in this case it can
serve as a young hot core tracer since its presence in this warm
region suggests it has recently been liberated from the icy mantles of
dust grains \citep[e.g.][]{Mangum1991}.  The \hcthreen\/ emission is
consistent with an upper temperature limit of \q 100 K; a more
quantitative determination is not possible with only two observed
transitions.  In contrast, the ratio of the \formald\/(\hhcoa) and
\formald(\hhcob) lines is a reliable density-independent temperature
diagnostic for T$_{K}\lesssim$ 50 K, and $N$(para-\formald)/$\Delta
v\lesssim$ 10$^{13.5}$ cm$^{-2}$ (\kms)$^{-1}$ \citep{Mangum93}. In
this regime, the \hhcoa/\hhcob ~ratio ranges from 15 to 5 for $T_k$=20
to 50 K. In contrast, the observed line ratios in S255N are less than
2.5 throughout the imaged region and are smallest ($\sim 1.5$) toward
SMA1, suggesting that the column density (i.e. opacity) and/or
temperature is too high for these lines to be diagnostic.  For the
position of SMA1-NE, assuming a temperature of 75 K, N$_{\rm
H_{2}CO}$\q 3.7 $\times$ 10$^{13}$ \cmsq from the \hhco\/(\hhcoa) line
and N$_{\rm H_{2}CO}$\q 1.0 $\times$ 10$^{14}$ \cmsq from the
\hhco\/(\hhcob). This comparison suggests that the \hhcoa\/ line is
moderately optically thick compared to the \hhcob\/ line, and that the
low line ratios are a combination of both the column density and
temperature being higher than the diagnostic range of these two
transitions. Combining this analysis with the SED models and the
single dish line results described above, the allowed temperature
range for SMA1 is 40 - 100 K. The resulting ranges of gas mass, column
density, and number density computed for SMA1 are shown in
Table~\ref{mass}.  The high derived gas density
(n$_{H_{2}}$\q3-16$\times$10$^{6}$~\cc), also implied by the presence
of the water maser, indicates that the gas and dust temperatures are
likely to be well-coupled \citep[e.g.][]{Kaufman98,Ceccarelli96}.
 
Unlike SMA1, SMA2 and SMA3 are not accompanied by significant line
emission.  \formald, \methanol, and SiO emission are present to the
north of SMA2, and CN emission is detected toward SMA3 (\S3.3), but
the physical relationship (if any) between this line emission and the
dust continuum sources is unclear.  Also unlike SMA1, SMA2 and SMA3
are not associated with mid-IR emission in any IRAC band (\S3.4).
Instead, SMA2 and SMA3 appear to be cold, dark, young mm cores,
without evidence for current star formation.  On the basis of the lack
of line and mid-IR emission towards SMA2 and SMA3, we adopt a lower
temperature limit of 20~K and an upper temperature limit of 40~K for
these sources.  The corresponding range of masses, column densities,
and number densities for SMA2 and SMA3 are tabulated in
Table~\ref{mass}.  The mass of each (7-17 \msolar\/ for SMA2, 6-13
\msolar\/ for SMA3) is sufficient to form a low to intermediate mass
star.  No other cores are detected in the field to a $5\sigma$ upper
limit of $M < 3$\msolar\ (at $T=20$~K).

\subsection{Velocity Structure and Outflows}

\begin{figure}
\epsscale{0.5}
\plotone{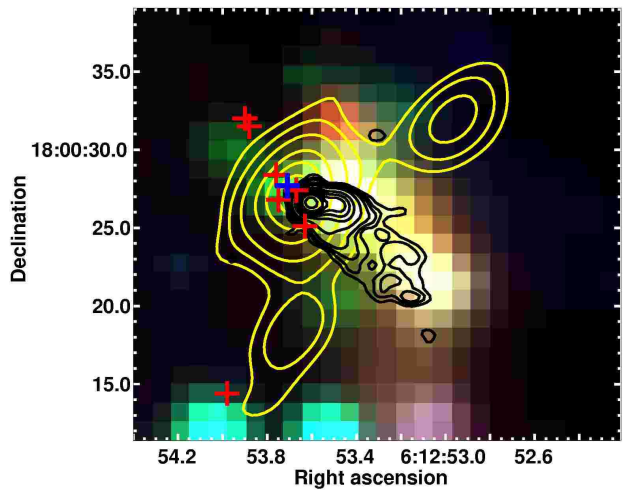}
\caption{Close-up three-color {\em Spitzer} IRAC image of S255N
  showing mid-IR emission offset to the NW of the UC \HII region with
  yellow SMA 1.3~mm and black 3.6~cm contours superposed.  The colorscale
  correspond to: 8.0 \um (red), 4.5~\um (green), and 3.6 \um (blue). The
  yellow SMA 1.3~mm continuum contour levels are (3, 7, 15, 31, 47,
  63)~$\times$~4 \mjb\/ (the rms noise).  The black VLA 3.6~cm continuum
  contour levels are (3, 5, 7, 11, 21, 41, 61, 81, 101, 121,
  161)~$\times$ 34 $\mu$Jy~beam$^{-1}$ (the rms noise).  Positions of
  Class I \methanol\/ masers from \citet{Kurtz04} are marked with red
  crosses, and the position of the newly-reported water maser is
  marked with a blue cross.}
\label{irzoom}
\end{figure}

Figure~\ref{irzoom} shows a close-up view of the {\em Spitzer} 3-color
IRAC image shown in Figure~\ref{irbig}.  The brightest mid-IR emission
is extended along a NE-SW axis, approximately parallel to the axis of
the UC\HII region, but with an offset to the northwest of $\sim
2\arcsec$.  The overall morphology of S255N is consistent with the
multi-band bright mid-IR emission tracing the surface of the UC\HII\/
region, which is less dense to the southwest (of SMA1), as indicated
by the diffuse ``tail'' of the cometary UC \HII region extending in
this direction.  However, the detailed interpretation of the mid-IR
emission toward S255N is complicated by the offset described above,
and the sharp cutoff of the 3.6 and 8.0 $\mu$m emission along the
southeast boundary of the UC\HII\/ region. Indeed, mid-IR emission is
notably absent toward SMA2 and SMA3, as well as toward much of SMA1. A
likely scenario for this behavior is absorption of the mid-IR emission
by the high column density mm cores; in this scenario the bulk of the
relatively cold mm cores must be in front of the UC\HII\/ region.

With the exception of SMA1, the molecular emission in S255N is not
obviously associated with any continuum emission, and is therefore
unlikely to be centrally heated. However, as described in \S 4.1 the
\formald\/ line ratios suggest the gas is warm. Thus, it is likely
that much of this emission is associated with outflow material,
although the number of outflows and their driving source(s) are
unclear.  Published data on large-scale outflows in the region
\citep[e.g.][]{Miralles97,Richardson85,Heyer89,Ruiz92} are
unfortunately too low in angular resolution to be useful in
distinguishing outflows associated with S255N from those associated
with S255IR to the south, and/or the maps are swamped by emission from
a large outflow flowing north from S255IR.  Excluding the ``SW''
position, the relatively narrow linewidths of these S255N line
emission regions suggest that they are density enhancements within a
larger extended flow resolved out by the interferometer. The relative
similarity of the line center velocities further suggests that the
outflows are mostly in the plane of the sky. 

The linear morphology of the (green) 4.5~\um emission northeast of
SMA1 is suggestive of an outflow (Fig.~\ref{irzoom}).  Such 4.5~\um
nebulosity is a conspicuous feature of IRAC images of star forming
regions. Recent analysis of the massive DR21 outflow, the best-studied
example, has shown that \h line emission accounts for \q50\% of the
observed 4.5 \um IRAC flux, and that the outflow morphology is almost
identical in IRAC 4.5~\um and narrow-band 2.122 \um (\h 1-0~S(1) line)
images \citep{Davis07, Smith06}. In S255N, the 2.122 \um \h clump
S255:H2-3 lies at the base of the 4.5~\um nebulosity; the \h clump is
also coincident, within reported astrometric uncertainties, with our
newly-reported water maser.  Both the 2.122 \um \h 1-0~S(1) line and
the 4.6947 \um \h 0-0 S(9) line, identified by \citet{Smith06} as the
dominant contributor to IRAC band 2, trace moderate-velocity shocks
\citep{Draine83}.  Recent models by \citet{SmithRosen} of shocks in
dense protostellar molecular jets predict that the integrated \h line
contribution to IRAC band 2 will be 5-14 times greater than to IRAC
band 1 (3.6 $\mu$m), consistent with the ratio of the emission seen in
these bands toward the linear 4.5 \um\/ feature.  The 44 GHz Class I
methanol masers, five of which lie along the 4.5~\um emission feature,
provide further evidence for its identification as an outflow.
\citet{Kurtz04} found that masers of this type are often found in
association with such outflow tracers as SiO (IRAS 20126+4104,
G31.41+0.31 and G34.26+0.15) and \h (IRAS 20126+4104).  In
G31.41+0.31, the 44 GHz methanol masers are also associated with
thermal methanol emission \citep{Kurtz04}, which \citet{LW97} found
traced shock/clump interfaces in the DR21 outflow.  The parallels
between these examples and S255N strongly suggest that the 44 GHz
methanol masers, \formald\/, \methanol\/, SiO, H$_{2}$, and 4.5~\um
emission extending northeast from SMA1-NE trace a molecular outflow
from a protostar, probably \smane\/.

At the SW position, the broad lines with strong blue wings combined
with the morphology of the \formald\/ and \methanol\/ channel maps are
consistent with a blueshifted outflow lobe driven by SMA1-SW.
Notably, no 44 GHz methanol masers coincide with the very strong
thermal \methanol\/ emission of the SW line peak, although elsewhere
in S255N the methanol maser flux densities are loosely correlated with
the strength of thermal \methanol\/ emission.  The absence of masers
towards the SW position suggests that the physical conditions are not
appropriate for the collisional pumping of Class I methanol masers
\citep{Cragg92, Plambeck90}.

\subsection{The Nature of SMA1}

\begin{figure}
\epsscale{0.5}
\plotone{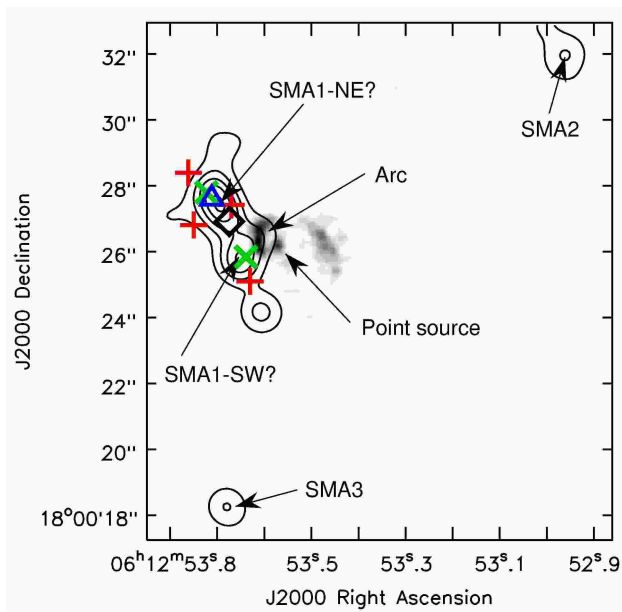}
\caption{Greyscale image of the high resolution VLA 3.6~cm emission
shown in Figure~\ref{masers}b with SMA 1.3 mm uniform weighted
continuum contours superposed. The 1.3 mm image was restored with a
$1\arcsec$ beam ($\sim 3$ times smaller than the longest observed
baseline) which essentially shows regions where clean components are
concentrated.  Red $+$ symbols show the 44 GHz methanol maser
positions, the blue $\triangle$ shows the H$_2$O maser position, the
green $\times$ symbols show the peak locations of the $\sim 7$
(SMA1-NE) and $\sim 11.5$ \kms\/ (SMA1-SW) H$_2$CO components, and the
black $\diamond$ shows the location of SMA1 reported in Table~\ref{cont}.  The
localization of clean components into two distinct regions in the
vicinity of SMA1 suggests the presence of at least two continuum
sources; this result requires confirmation by higher resolution
mm/submm data.}
\label{super}
\end{figure}

The complex kinematic behavior of the molecular line emission in the
vicinity of SMA1 including SMA1-NE, SMA1-SW, and position ``SW''
(\S3.3) is difficult to explain in the context of a single protostar.
Though SMA1-NE and SMA1-SW could be interpreted as the blue and
red-shifted lobes, respectively, of an outflow, this scenario does not
explain the very blue-shifted emission further to the southwest at
position ``SW''. Rotation also seems like an unlikely explanation for
the velocity gradient between SMA1-NE and SMA1-SW since the gradient
is parallel to the direction of the two probable outflow regions: the
4.5 \um emission to the northeast and ``SW'' to the southwest.
Instead, the combination of the chemical and kinematic differentiation
between SMA1-SW and SMA1-NE suggests the presence of two individual
sources, one at +7 \kms\ and one at +11.5 \kms.  To investigate this
possibility, in Figure~\ref{super} we show a uniform weighted SMA
millimeter continuum image restored with a beam of 1\farcs0 ($\sim 3$
times smaller than the longest observed baseline), which essentially
reveals the location of the clean components.  The localization of
clean components into two main regions in the vicinity of SMA1
suggests the presence of at least two sources separated by \q
1\farcs84 (4800~AU).  If these two clean component peaks correspond to
real dust sources, their positions are in good agreement with the two
kinematically distinct formaldehyde peaks ($<0\farcs1$ and
$<0\farcs5$).  The northeast component of the pair is also within
0\farcs2 of the water maser. That two distinct protostars would exist
with this separation is reasonable, as multiplicity of protostars has
been observed on scales of $<$6,000 AU
\citep[e.g.][]{Megeath05}. Although the presence of two protostars is
a plausible interpretation, it clearly requires higher resolution
continuum observations for confirmation.

In any case, the molecular line emission from SMA1-NE and SMA1-SW is
reminiscent of hot molecular cores (HMCs),
particularly the detection of DCN and \hcccn.  These molecules--like
\methanol\/ and \formald\/, which also show strong emission towards
\smane\ and \smasw\--are present in the gas phase in HMCs because they
have been evaporated from grain mantles
\citep[e.g.][]{Caselli05,Sz05}.  Complex organic molecules such as
HCOOCH$_{3}$, however, are believed to be ``daughter'' species, formed
in the gas phase by reactions of ``parent'' species such as \formald\/
and \methanol\/ \citep{Caselli05}.  Thus, while \smane\ and \smasw\ do
not exhibit the truly copious molecular emission observed towards some
HMCs \citep[e.g.][]{Hatchell98,Schilke06}, this is consistent with
\smane\ and \smasw\ being very young sources, in which gas-phase
hot-core chemistry has not yet produced abundant complex organic
molecules.

The line emission from the SMA1 sources is unusual in that the
DCN(3-2) emission is stronger than \hcccn\/(\hcccnlineanoarr\/).  By
modeling deuterium chemistry, \citet{RobertsMillar00} find that the
steady-state abundance of DCN in molecular clouds is a complicated
function of temperature and density (see their Figure 7), but
generally higher at low metallicity.  We note this because S255N is
located approximately ($l \sim 192^\circ$) in the direction of the
Galactic anticenter, and may have lower metallicity than inner-galaxy
star-forming regions \citep{Daflon2004,Afflerbach1997}.  Our SMA data,
however, do not allow us to disentangle the effects of abundance and
excitation on the strengths of DCN and \hcccn\/ emission.

The geometry of a HMC located a few arcseconds ahead of the vertex of
a cometary UC\HII region has been seen in other objects observed at
high angular resolution, and this notable configuration has led to
much discussion on the energy source responsible for these HMCs.  The
best-studied case for external heating is G34.26+0.15, in which
\citet{Watt99} and \citet{Mookerjea07} argue that HMC emission
(characterized by complex nitrogen and oxygen-rich molecules) arises
in gas heated by component C, the most evolved of three nearby UC\HII
regions.  In the absence of extinction, a 1.1 $\times$ 10$^{4}$ \lsun~
source at the location of the 3.6~cm point source could heat \smane\/
to \q37 K, and \smasw\/ to \q 51 K, consistent only with the lower end
of the range of plausible gas temperatures.  In contrast, G29.96-0.02
is the prototype for a HMC located ahead of a cometary UC\HII region
and internally heated by a high mass protostellar object
\citep{Buizer05,Gibb03}.  In G29.96-0.02, HMC emission is coincident
with a resolved 1.4 mm continuum source, water maser spots, and a
mid-IR sub-arcsecond point source \citep[][and references
therein]{Buizer05, Gibb03, Olmi03}.  In S255N, the arrangement of the
1.3~mm and 3.6~cm sources along with the presence of water maser
emission coincident with the molecular emission closely resembles the
case of G29.96-0.02.  We favor the interpretation that one or more
sources younger than the excitation source of the UC\HII region are
present and responsible for the compact dust and molecular line
emission from SMA1, consistent with the interpretation of the hot core
emission outlined above.

\section{Conclusions}

Our multiwavelength observations of S255N reveal significant new
details in this luminous star-forming region.  While the
previously-identified UC\HII\ region dominates the cm continuum and
mid-IR emission, the 1.3~mm continuum emission has been resolved into
three compact cores (SMA1, SMA2, and SMA3) clustered on scales of
0.1-0.2~pc.  Dominated by dust emission, these cores range in mass
from 6 to 35 M$_\sun$.  There are no mid-infrared point source
counterparts to any of the dust cores, suggesting an early
evolutionary phase.  The spectral line emission at the position of the
brightest core, SMA1, is spatially compact and includes HC$_3$N, CN,
DCN, CH$_3$OH, SiO, and H$_2$CO.  SMA1 appears to be a developing hot
core offset by a few thousand AU from the UC\HII\ region.  The
chemical and kinematic structure toward SMA1 is suggestive of further
multiplicity at these scales.  A 4.5 $\mu$m linear feature emanating
to the northeast of SMA1 is aligned with a cluster of methanol masers
and likely traces a outflow from a protostar within SMA1.  We conclude
that S255N is actively forming a cluster of intermediate to high-mass
stars.  In addition, we speculate that some of the missing flux in the
SMA continuum image could be in the form of additional compact
low-mass, cold dust cores that lie below the sensitivity limit of our
observations ($M \sim 3$\msolar\ at $T=20$~K).  Higher-resolution and
more sensitive observations are needed to search for additional
protostars in S255N and other young protoclusters. Resolving
individual protostars in regions like these is a necessary task to
determine how dense these young protoclusters are, and how
interactions among protostars in protoclusters may affect the process
of star formation.

\acknowledgments

This work is based in part on observations made with the {\it Spitzer
Space Telescope}, which is operated by the Jet Propulsion Laboratory,
California Institute of Technology under a contract with NASA.  This
research has made use of NASA's Astrophysics Data System Bibliographic
Services and the SIMBAD database operated at CDS, Strasbourg, France.
Research at the CSO is funded by the NSF under contract AST96-15025.
CJC is supported by a National Science Foundation Graduate Research
Fellowship and acknowledges partial support from a Wisconsin Space
Grant Graduate Fellowship. CJC would like to thank the SMA and NRAO
for student research support.

\clearpage

\begin{deluxetable}{cccccccc}
\tablewidth{0pc}
\tablecaption{Properties of millimeter continuum sources in S255N \label{cont}}  
\tablecolumns{8}
\tablehead{
\colhead{Source} & \multicolumn{2}{c}{J2000 coordinates} &
\colhead{$I_{\rm 1.3mm}$} &
\colhead{Size\tablenotemark{a}} & \colhead{$F_{\rm 1.3mm}$\tablenotemark{b}} &
\colhead{$T_{\rm b}$\tablenotemark{c}}\\ 
\tableline
\colhead{}       &  \colhead{$\alpha$ ($^{\rm h}~~^{\rm m}~~^{\rm
    s}$)}   & \colhead{$\delta$ ($^{\circ}~~{'}~~{''}$)} &
\colhead{(Jy/b)} & \colhead{[~\pp~$\times$~\pp~(\arcdeg)]} & \colhead{(Jy)}& \colhead{(K)}} 
\startdata
SMA1 & 06 12 53.67 & +18 00 26.9  & 0.29  & 3.9$\times$2.0 (17.4) & $0.58 \pm 0.12$ & 1.84\\ 
SMA2 & 06 12 52.97 & +18 00 31.9  & 0.09  & 2.2$\times$1.6 (97.5) & $0.12\pm 0.02$ & 0.82 \\ 
SMA3 & 06 12 53.69 & +18 00 18.5  & 0.05  & 5.2$\times<$1.3 (169.6) & $0.09 \pm 0.02$ & 0.33 \\   
\tableline
Total &              &            &         & & $0.79 \pm 0.16$\\
\enddata
\tablenotetext{a}{Deconvolved source size determined by fitting a
  single Gaussian component to each source.  The SMA beam is $4\farcs7
  \times 2\farcs4$ (P.A.=$-45.65\arcdeg$).}
\tablenotetext{b}{Uncertainties include 20\% calibration uncertainty.}
\tablenotetext{c}{Brightness temperature computed using the
  Rayleigh-Jeans approximation.}
\end{deluxetable}

\clearpage
\begin{deluxetable}{llcc}
\tablewidth{0pc}
\tablecolumns{4}
\tablecaption{Molecular species and transitions observed in S255N \label{trans}}
\tablehead{
\colhead{Species} & \colhead{Transition} & \colhead{Frequency}  & \colhead{E$_{\rm upper}$/k} \\
\colhead{}  & \colhead{} &   \colhead{(GHz)}    &  \colhead{(K)}
}    
\startdata
SiO      & \sioline         & 217.104980  & 31.2 \\
DCN      & \dcnline         & 217.238538  & 20.9 \\
\hhco  & \hhcolinea         & 218.222192  & 21.0 \\
\hcccn   & \hcccnlinea      & 218.324723  & 131  \\
\chhhoh  & \chhhohline      & 218.440050  & 45.5 \\ 
\hhco  & \hhcolineb         & 218.475632  & 68.1 \\
\hhco  & \hhcolinec         & 218.760066  & 68.1 \\
CN\tablenotemark{a,b} & \cna  & 226.874191  & 16.4 \\
CN\tablenotemark{a,c} & \cnb  & 226.874781  & 16.4 \\
CN\tablenotemark{a} & \cnc  & 226.875896  & 16.4 \\
CN                  & \cnd                   & 226.887420  & 16.4 \\
CN                  & \cne                   & 226.892128  & 16.4 \\
\hcccn   & \hcccnlineb      & 227.418905  & 142  \\
\enddata
\tablenotetext{a}{These components are blended in our spectra.}
\tablenotetext{b}{The CN hyperfine components at frequencies lower than this
     one lie outside of our observed bandpass.}
\tablenotetext{c}{This transition is used to set the velocity scale for CN in Fig~\ref{cpro}.}
\end{deluxetable}

\clearpage

\begin{deluxetable}{ccccccccccccc}
\tablewidth{0pc}
\tablecolumns{3}
\tablecaption{Spectral line positions \label{linepos}}
\tablehead{
               & \multicolumn{2}{c}{J2000 coordinates} \\
\colhead{Name} & \colhead{$\alpha$ ($^{\rm h}~~^{\rm m}~~^{\rm s}$)}  &  
                 \colhead{$\delta$ ($^{\circ}~~{'}~~{''}$)} 
}
\startdata
SMA1-NE & 06 12 53.73 & +18 00 27.8\\
SMA1-SW & 06 12 53.64 & +18 00 25.4\\
SW      & 06 12 53.45 & +18 00 23.0\\
NE      & 06 12 53.76 & +18 00 33.2\\
N       & 06 12 53.70 & +18 00 39.8\\
NW      & 06 12 52.86 & +18 00 36.2\\
\enddata
\end{deluxetable}

\begin{deluxetable}{ccccccccccccc}
\rotate
\tablewidth{0pc}
\tabletypesize{\scriptsize}
\tablecolumns{13}
\tablecaption{Fitted line properties\label{linefits}}
\tablehead{
\colhead{} & \multicolumn{3}{c}{\formald(\hhcolineanoarr)} & \multicolumn{3}{c}{\methanol} & 
\multicolumn{3}{c}{DCN} & \multicolumn{3}{c}{\hcccn\/(24-23)}\\
\tableline
\colhead{Position} & \colhead{Center} & \colhead{Width} & \colhead{$\int S_vdv$} & \colhead{Center} & \colhead{Width} & \colhead{$\int S_vdv$} 
& \colhead{Center} & \colhead{Width} & \colhead{$\int S_vdv$} & \colhead{Center} & \colhead{Width} & \colhead{$\int S_vdv$}\\
\colhead{} & \colhead{\kms} & \colhead{\kms} & \colhead{\jyb*\kms} & \colhead{\kms} & \colhead{\kms} & \colhead{\jyb*\kms} 
& \colhead{\kms} & \colhead{\kms} & \colhead{\jyb*\kms} & \colhead{\kms} & \colhead{\kms} & \colhead{\jyb*\kms}}
\startdata
\smane\/ & 6.9(0.3)\tablenotemark{a} & 5.9(0.6)\tablenotemark{a} & 5.5(0.7)\tablenotemark{a} & 6.6(0.3)\tablenotemark{c} & 3.4(0.9)\tablenotemark{c} & 0.9(0.3)\tablenotemark{c} 
& 8.2(0.1) & 3.3(0.2) & 5.2(0.4) & 6.3(0.3)\tablenotemark{a} & 3.5(0.8)\tablenotemark{a} & 1.9(0.6)\tablenotemark{a}\\
\smasw\/ & 12.1(0.1) & 2.9(0.2) & 4.2(0.3) & 11.2(0.1) & 2.4(0.2) & 3.1(0.4) 
& 10.7(0.3)\tablenotemark{b} & 5.6(0.7)\tablenotemark{b} & 3.1(0.5)\tablenotemark{b} & 9.5(0.8)\tablenotemark{a} & 11.2(2.1)\tablenotemark{a} & 3.0(0.7)\tablenotemark{a} \\
SW & 6.3(0.1) & 9.2(0.4) & 24.3(1.2) & 7.8(0.1) & 6.9(0.3) & 13.4(0.8) & \nodata & \nodata & \nodata & \nodata & \nodata & \nodata \\
NE & 8.3(0.1) & 2.9(0.3) & 2.4(0.4) & 8.3(0.2) & 2.5(0.4) & 1.6(0.4) & \nodata & \nodata & \nodata & \nodata & \nodata & \nodata \\
N & 11.5(0.1) & 2.6(0.3) & 2.4(0.3) & 10.1(0.2) & 2.5(0.5) & 1.5(0.4)& \nodata & \nodata & \nodata & \nodata & \nodata & \nodata \\
NW & 8.9(0.2) & 5.8(0.6) & 5.3(0.7) & 8.3(0.2)\tablenotemark{a} & 5.7(0.5)\tablenotemark{a} & 3.7(0.5)\tablenotemark{a}& \nodata & \nodata & \nodata & \nodata & \nodata & \nodata \\
\enddata
\tablenotetext{a}{Not well fit by a single Gaussian.}
\tablenotetext{b}{Gaussian fit encompasses two blended components.}
\tablenotetext{c}{Parameters for second-strongest velocity component.  Strongest component is very similar to that towards \smasw\/.}
\end{deluxetable}

\clearpage

\begin{deluxetable}{ccccccc}
\tablewidth{0pc}
\tablecolumns{7}
\tablecaption{Range of estimated masses of dust cores in S255N \label{mass}}  
\tablehead{
 \colhead{}      & \colhead{$\kappa$} & \colhead{$T_{dust}$}  & \colhead{$\tau_{dust}$} & \colhead{$M$} & \colhead{$N_{\rm H_{2}}$\tablenotemark{b}} & \colhead{$n_{\rm H_{2}}$\tablenotemark{b}}       \\
\colhead{Source\tablenotemark{a}} & \colhead{(cm$^2$~g$^{-1}$)} &
\colhead{(K)} & \colhead{(1.3mm)} & \colhead{($M_\odot$)} &
\colhead{(10$^{23}$cm$^{-2}$)}& \colhead{(10$^{6}$cm$^{-3}$)}} 
\startdata
SMA1\tablenotemark{c} & 1 & 40-100 & 0.04-0.02 & 35-13 & 10.5-3.8 & 15.9-5.8\\
SMA2                  & 1 & 20-40 & 0.02-0.01 & 17-7 & 5.0-2.2 & 7.6-3.3 \\
SMA3                  & 1 & 20-40 & 0.01-0.01 & 13-6  & 3.9-1.7 & 5.9-2.5 \\
Total                 &   &       &           & 65-26 &  &  \\ 
\enddata
\tablenotetext{a}{Assumed distance is 2.6~kpc.}
\tablenotetext{b}{Beam-averaged quantities.The SMA beam is
$4\farcs7\times2\farcs4$ (P.A.=$-45.65\arcdeg$).}
\tablenotetext{c}{The mass of SMA1 was calculated using the 1.3~mm flux density less 
the estimated free-free contribution of 20~mJy.}
\end{deluxetable}

\clearpage

\clearpage

\clearpage

\clearpage

\end{document}